\documentclass[runningheads]{llncs}
\usepackage{graphicx}
\usepackage{caption}
\usepackage{multirow}
\usepackage{subfigure}
\usepackage{comment}
\usepackage{amsmath,amssymb} 
\usepackage{color}

\usepackage[width=122mm,left=12mm,paperwidth=146mm,height=193mm,top=12mm,paperheight=217mm]{geometry}

\begin{document}
\pagestyle{headings}
\mainmatter
\def\ECCVSubNumber{4633}  

\title{Concatenated  Attention Neural  Network for Image Restoration} 


%
\author{YingJie Tian\inst{1,2,3}\and
YiQi Wang\inst{2,3,4} \and
LinRui Yang\inst{2,3,5} \and ZhiQuan Qi\inst{1,2,3}}
\authorrunning{Tian et al.}
%
\institute{School of Economics and Management, University of Chinese Academy of Sciences, Beijing, China \and
Research Center on Fictitious Economy and Data Science, Chinese Academy of Sciences, Beijing, China\\
\and
Key Laboratory of Big Data Mining and Knowledge Management Chinese Academy of Sciences, Beijing, China\\
\and School of Computer Science Technology, University of Chinese Academy of Sciences, Beijing, China\\
\and Sino-Danish College, University of Chinese Academy of Sciences, China\\
\email{wangyiqi18@mails.ucas.edu.cn}
}
\maketitle

\begin{abstract}
In  this  paper,  we  present  a general framework for low-level vision tasks including image compression artifacts reduction and image denoising. Under this framework, a novel concatenated attention neural network (CANet) is specifically designed for image restoration. The main contributions of this paper are as follows: First, by applying concise but effective concatenation and feature selection mechanism, we establish a novel connection  mechanism  which connect different modules in the modules stacking network. Second, both pixel-wise and channel-wise attention mechanisms are used in each module convolution layer, which promotes further extraction of more essential information in images. Lastly, we demonstrate that CANet achieves better results than previous state-of-the-art approaches with sufficient experiments in compression artifacts removing and image denoising.

\end{abstract}

\section{Introduction}

\begin{figure}
\setlength{\belowcaptionskip}{-1cm} 
\centering
\subfigure[original]{
\includegraphics[width=1.5in]{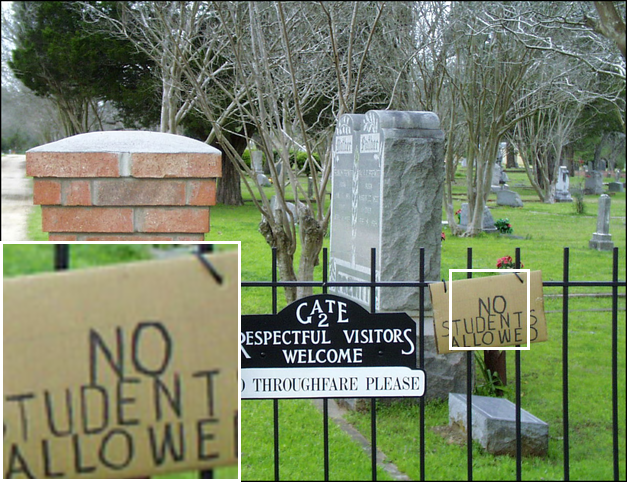}
}	
\subfigure[jpeg]{
\includegraphics[width=1.5 in]{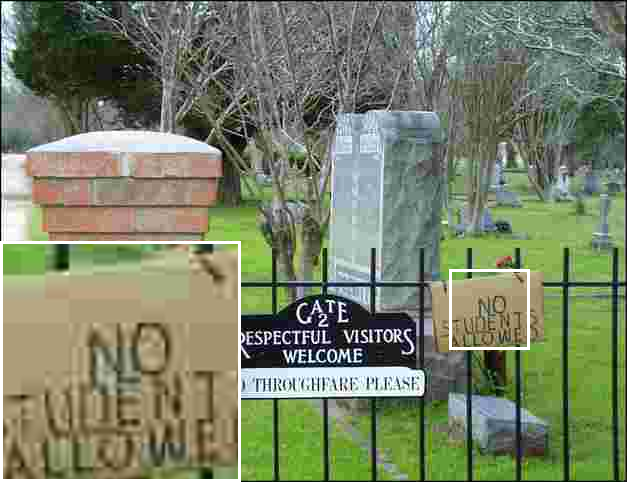}
}
\quad
\subfigure[ARCNN]{
\includegraphics[width=1.5 in]{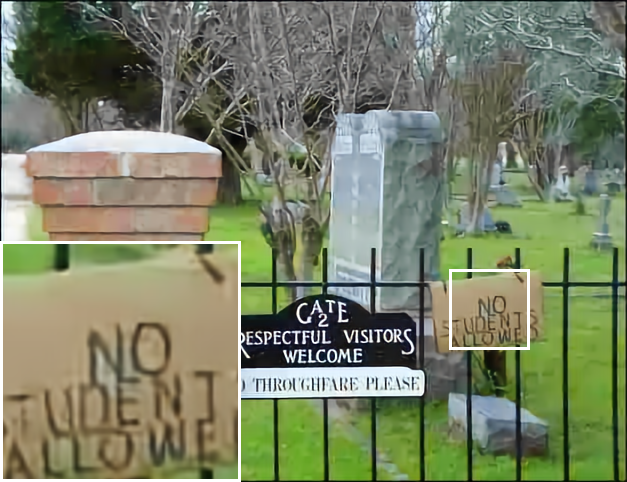}
}	
\subfigure[GLCNet]{
\includegraphics[width=1.5in]{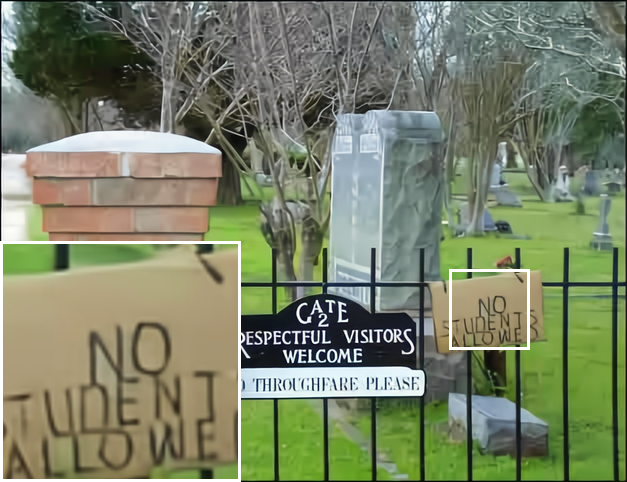}
}
\caption{Visual example of image compression artifacts reduction. Compression quality=10 'cemetry' from LIVE1.}
\label{first fig}
\end{figure}

Image restoration(IR) aims at recovering high-quality(HQ) images from corresponding corrupted low-quality(LQ) images. It is an ill-posed inverse procedure, which has always been a challenge but popular research area in computer vision field. In plenty of application scenarios, IR can achieve a good visual experience for users and assist high-level vision applications.  Fig.\ref{first fig} shows a visual example for image compression artifact reduction. 

With the popularity of deep learning, IR also ushered in the spring.  In 2016, Dong et al first proposed a  compression artifact reduction method based on deep convolutional network, and got  the best performance at the time  easily\cite{Dong2016Compression}. Later,  Zhang et al took  full advantage of the pixel  and  discrete cosine transform(DCT) domain  and employed multi-scale convolu-tional neural network to remove the JPEG compression artifacts\cite{DBLP:journals/corr/abs-1806-03275}. And \cite{DBLP:journals/corr/abs-1810-08042} proposed novel implicit  dual-domain  convolutional  network  for  robust  color  image  compression  artifact  reduction.  Zhang et al proposed new DnCNN method\cite{zhang2017beyond}  to deal with the issue  by introducing residual learning and batch normalization.  \cite{tai2017memnet} designed a new memory network called MemNet for image restoration.  In \cite{guo2019toward},  a new noise estimation subnetwork was designed  for further improving  the efficiency of the model for image denosing. \cite{anwar2019real} proposed  the channel attention to remove the noise high efficiency. In addition, Zhang et al introdued a very deep network with the non-local attention, and achieved  promising results in both image denoising and compression artifacts reduction tasks\cite{zhang2019residual}. 

However, there are still some issues in the methods that illustrated before.
\textbf{First}, many algorithms use a huge amount of complicated blocks to constitute their network, and single element-wise operation for connecting blocks leads to the inefficiency of extracting information.
\textbf{Second}, most algorithms did not explore the attention mechanism to perform the task\cite{show2015tell}, or few ones used it only under specific modules. Thus, attention mechanism has not been taken throughly advantage in IR prior  research. On the other hand, in order to minimize bad influence of the noise and corrupted information, IR is a typical task that requires employment of image information with caution. If the attention network is only added to the local part, the unfiltered noise will interfere final results within feature maps in the subsequent process. In conclusion, a better attention network needs to be developed to solve extremely corrupted images.

To address above issues, this paper proposes a general framework for many low-level vision tasks. The concatenation between blocks and attention mechanism in blocks are employed for image restoration under this framework, which accelerates the formulation of concatenated attention neural network  (CANet). In CANet, each block (called attention block(A-block)), is made up of residual modules based on pixel-wise attention and channel-wise attention mechanism. Notice that, in one feature map, some of its pixels and channels are likely to contain more noisy information than others. A-block can reduce the bad effects brought by these information in both pixel-wise and channel-wise. At the same time, among each attention block(A-block) which is composed by multiple layers, the fashion of  conveying the output of each block directly to following ones is abandoned. A new feature reuse mechanism is introduced. It can fuse current output to the next block with the features extracted from the previous blocks and go through a feature selection operation in local and global way. By doing this, final features can be mixed both locally and globally. Next, feature reuse operation supervises the output of feature information. At last, instead of ResNet\cite{he2016deep} element-wise operation, the concatenation operation is employed to avoid the loss of valid information. Experiments demonstrate that CANet achieves better results than previous state-of-the-art approaches in compression artifacts removing and image denoising tasks.

	The main contributions of this work are three-fold:
	
	1) A new general framework for image restoration is proposed, which uses a modules stacking method to achieve feature reuse and feature selection. Concatenation and feature selection operation are used to reuse the features both locally and globally. The process of feature fusion in different modules can make the network recovering the image quality in an evolutionary way.
	
	2) We propose a quite lightweight attention residual module, consisting of pixel-wise attention and channel-wise attention. The pixel-wise attention is used to filter pixel spatial information, while the channel-wise module is used to collect the overall information on channel-wise and then we perform channel attention mechanism. All of these make it possible to get effective information from the hierarchical features.
	
	3)  We demonstrate CANet achieves a better performance than some state-of-the-art methods among various image restoration tasks like image denoising, compressed artifacts reduction.

\section{Related Work}
\subsection{Image Restoration Neural Network.}
In view of some successful deep learning methods, including SRCNN\cite{dong2014learning} for super resolution task and a four-layer CNN\cite{Dong2016Compression} for image compress artifacts reduction, in low-level vision tasks, how to get a high performance network architecture has always been a hot topic. For instances, Zhang et al proposed a deep network named DnCNN\cite{zhang2017beyond} for image denosing based on residual learning and batch normalization\cite{ioffe2015batch}, after which, the introduction of residual learning made a great progress in image denoising. Lim et al\cite{lim2017enhanced} designed the EDSR by removing the batch normalization layer, and achieved a new state-of-the-art performance at that time. Huang et al\cite{huang2017densely} proposed DenseNet using the dense connection among convolution components. Instead of blindly pursuing deeper and wider neural networks, they explored the measure to more efficiently reuse features. Qiu et al\cite{qiu2019embedded} proposed EBRN, a recursive network for single-image super resolution, which made the shallow and deep layer to recover the low-frequent  and  high-frequent information of images, respectively. 

\subsection{Attention Mechanism}
	Recently, there have been several innovative approaches related to attention mechanism. Hu et al\cite{hu2018squeeze} proposed squeeze-and-excitation(SE) block for image classification by using channel-wise attention mechanism. They creatively used the global pooling to collect global information, following with two convolution layers to compute the channel attention weight. Moreover, Zhang et al\cite{zhang2018image} designed the RCAN by introducing the channel-wise attention to solve the super-resolution problem. And RIDNet\cite{anwar2019real} combined the dilated convolution and channel-wise attention for image denoising and achieved a superior performance. Furthermore, Zhang et al\cite{zhang2019residual} proposed a deep residual non-local network(RNAN) for image restoration tasks. RNAN shows that non-local attention improves the representation ability of network, based on which, Qin et al\cite{qin2019ffa} proposed FFA-Net for single image dehazing by incorporating the pixel-wise and channel-wise attention mechanism, which greatly improved the network's performance.

\section{Concatenated Attention Network}
\subsection{Concatenation Framework}

\begin{figure}
\centering
\includegraphics[height=2.8cm]{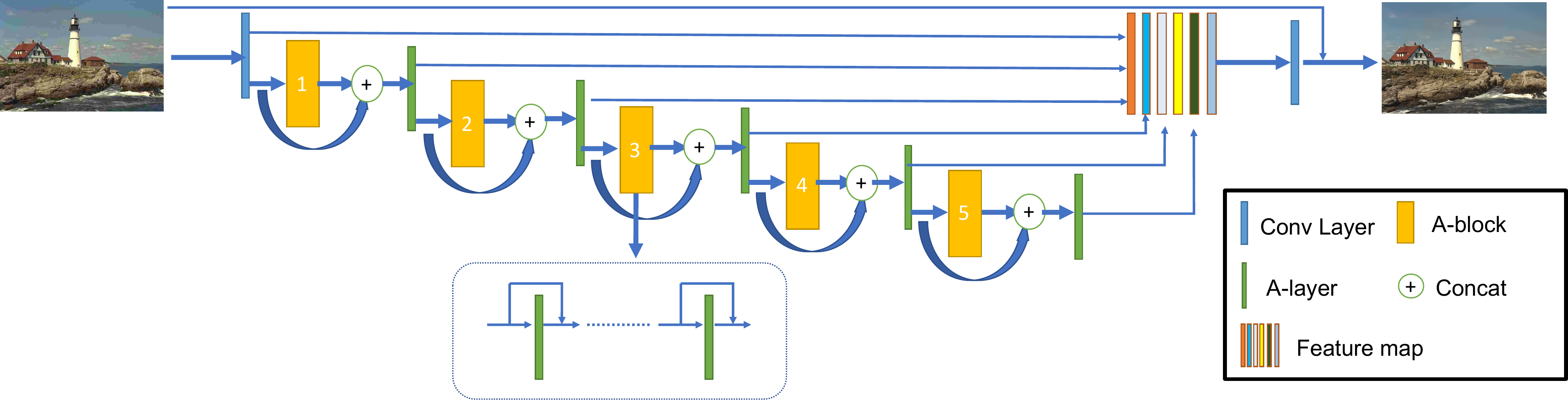}
\caption{The framework of our proposed concatenated attention residual network.}
\label{framework}
\end{figure}

In this section, we will illustrate the details of the proposed Concatenated Attention Network (CANet) model. The architecture is showed in Fig.\ref{framework}, which mainly consists of $N$ attention blocks (A-block) and skip connections.

As shown in Fig.\ref{framework}, CANet uses  A-blocks to step-wise reconstruct the high-quality images. The A-block consists of several A-layers, each of which is constituted by pixel-wise and channel-wise  attention layers. The transmission of features between two blocks is done by local and global skip connection. Here, Let's set $I_{LQ}$ for the low-quality images and $I_{HQ}$ for the high-quality images. The first layer extracts features $F_0$:
\begin{align}
F_0=H_{FE}\left(I_{LQ}\right),
\end{align}
where $H_{FE}()$ is a convolution layer  transforming images from pixel space to feature space. $I_{LQ}$ and $F_0$ are used for global residual learning by a long range skip connection. Then, we further have: 
\begin{align}
F_1=H_{A_1}\left(F_0\right),
\end{align}
where $H_{A_1}()$ denotes the first A-block. $F_1$ is also passed on to the global residual  learning. Therefore, we have:
\begin{align}
F_n=H_{A_n}\left(H_{Alayer}(concatenate(F_{n-1},F_{n-2}))\right), 
\end{align}
where $F_n$ is the output of n-th A-block(except $F_0$). Each output of blocks is passed to attention layer and has a local skip connection with the next block's output. $H_{Alayer}$ denotes the attention layer and will be  introduced in detail in subsequent chapters. $F_{n-1}$ and $F_{n-2}$ are concatenated for the local skip connection. 
In the end, we have 
\begin{align}
O_{F}=H_{FF}\left(concatenate(F_{0},F_{1}......F_{n})\right),
\end{align}
and 
\begin{align}
O_{HQ}=conv\left(O_{F}\right)+I_{LQ}.
\end{align}
Here, $F_{0}......F_{n}$ are all concatenated for global skip connection.  $H_{FF}$ is a composite function, and  $O_{HQ}$ is the final output image of the network. By using the global residual learning(GRL), the network actually learns the residual between HQ images and LQ images.

\subsection{Attention Residual Block}
  In this section, we will present details about Attention residual block(A-block). In many computer vision tasks, enormous experiments show that attention mechanism exhibits superior performance. However, most networks designed for compression artifact reduction and image denoising treat the pixel-wise and channel-wise equally, which probably limits the representational ability of network. In order to improve this, CANet employs special attention residual block constituted by pixel-wise and channel-wise attention layers(A-layer). As shown in Fig.2.  A-block is comprised by $N$ A-layers:
 \begin{align}
O_{n}=H_{n}(O_{n-1}+O_{n-2}),
\end{align}
where $O_{n}$ is the output of $n$-th A-layer and $H_{n}$ is the function of $n$-th A-layer. A-block is a recursive computation unit. The final output of A-block is $O_{N}$.


\begin{figure}
\setlength{\belowcaptionskip}{-1cm} 
\centering
\subfigure[SRResNet]{
\includegraphics[scale=0.5]{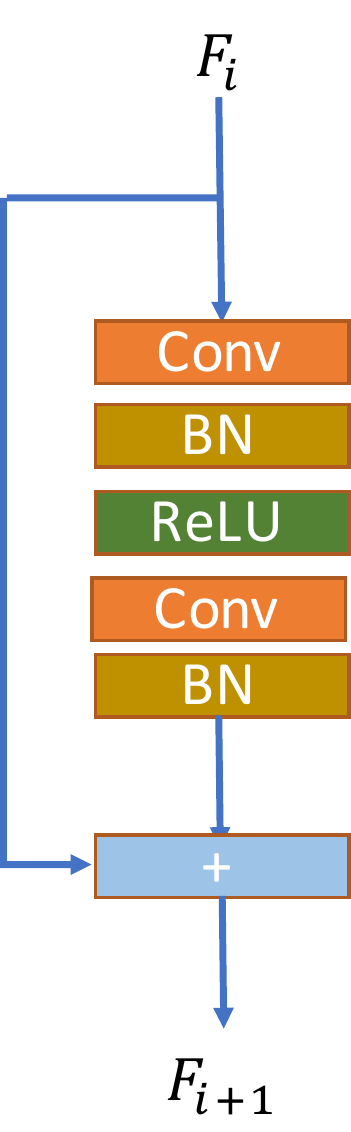}
}\hspace{6mm}	
\subfigure[EDSR]{
\includegraphics[scale=0.5]{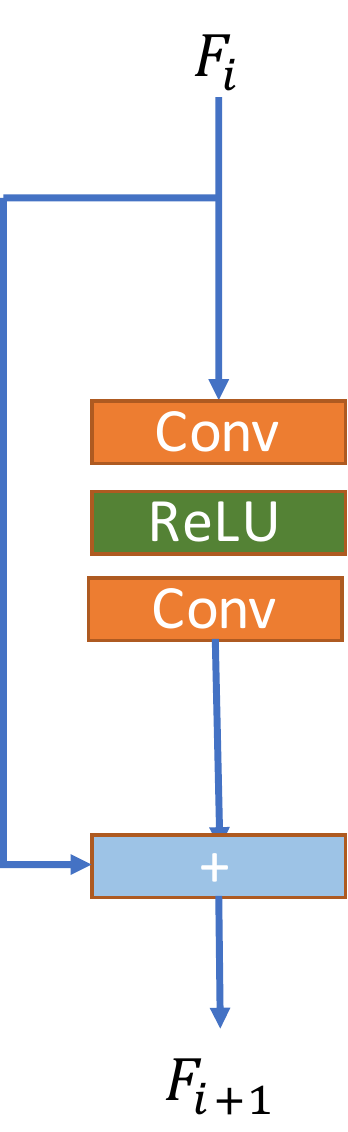}
}\hspace{6mm}	
\subfigure[RCAN]{
\includegraphics[scale=0.5]{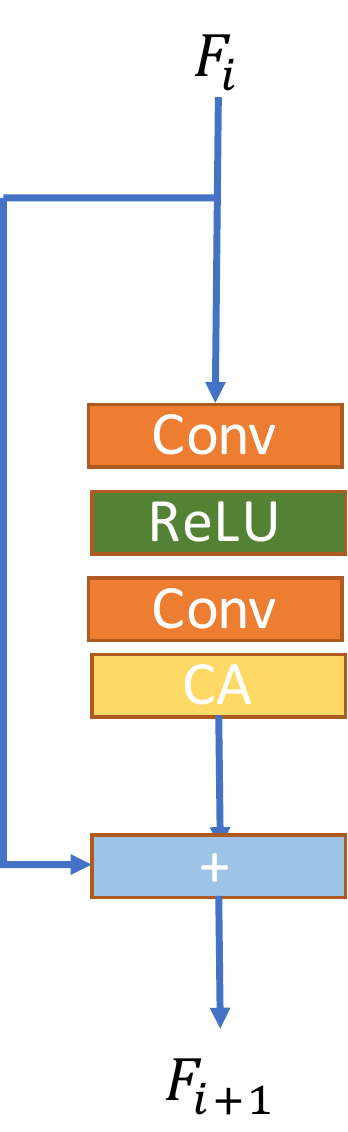}
}\hspace{6mm}		
\subfigure[CANet]{
\includegraphics[scale=0.5]{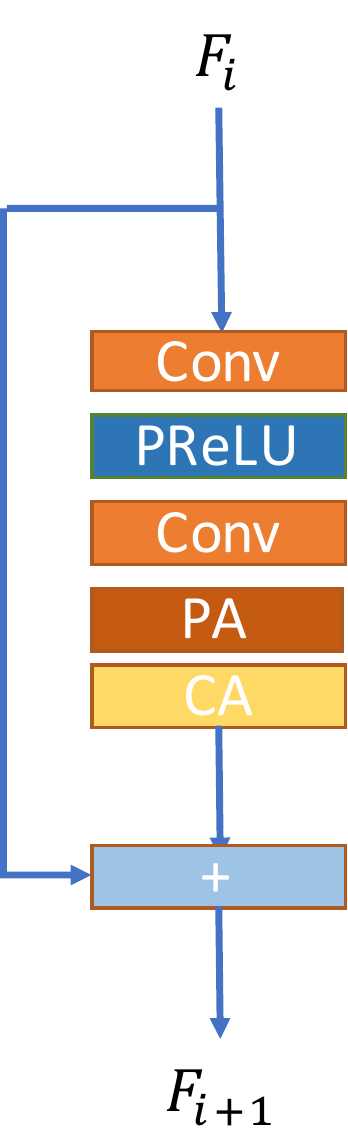}
}\hspace{6mm}		
\caption{Comparision of residual layers in SRResNet, EDSR, RCAN, and CANet.}	
\label{A-layer}
\end{figure}
                                                    
 In Fig.\ref{A-layer}, based on previous residual network research, by introducing the pixel-wise attention layer before the channel attention layer, the local residual mechanism is developed for deep training,  and then  multiple A-layers are used to form one A-block in CANet.
 
\begin{figure}
\centering
\subfigure[Pixel Attention Layer]{
\includegraphics[scale=0.5]{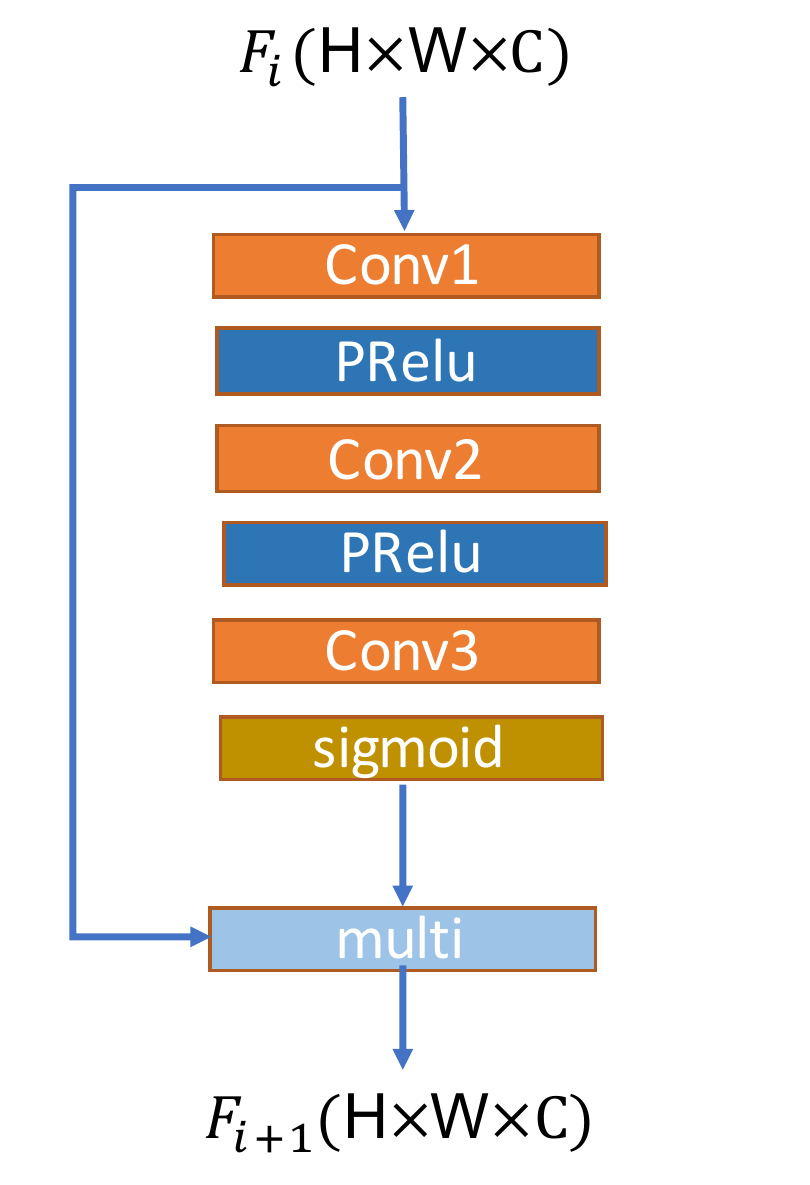}
}\hspace{10mm}	
\subfigure[Channel Attention Layer]{
\includegraphics[scale=0.5]{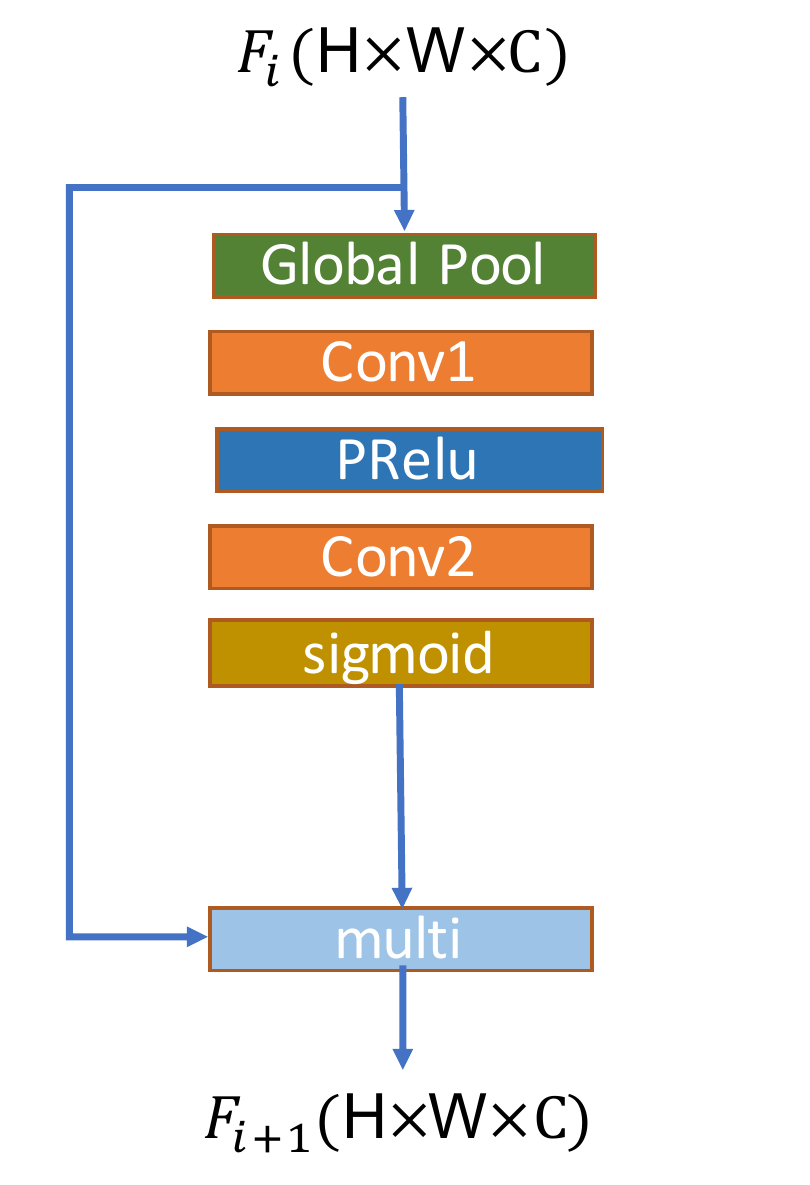}
}
\caption{The architecture of feature attention for pixel-wise and channel-wise.}	
\label{PAandCA}
\end{figure}

\subsection{Pixel-wise Attention}

Pixel-wise attention(PA) layer is showed as Fig.\ref{PAandCA}. CANet feeds the features from previous convolution layers into PA-layer, which is constituted by three $1\times1$ convolution layer, two PRelu  and one sigmoid activation function. The mechanism can be showed as
\begin{align}
r_p=\alpha(Conv_3\left (\delta_2 (Conv_2\left (\delta_1 (Conv_1\left(x_p\right))))\right)\right),
\end{align}
where  $Conv_1$,$Conv_2$ and $Conv_3$ are all $1\times1$ convolutions who  gradually reduce the dimension of the feature map channel from $H \times W \times C$ to $H \times W \times 1$. $\delta_1$ and $\delta_2$ are the PRelu activation functions and $\alpha$ is the sigmoid operator. And the output of sigmoid $r_p$ is the weighs of  features. Here, we rescale the $r_p$ for element-wise multiplying the features as 
\begin{align}
y_p = r_p \times x_p.
\end{align}
Through the structure of A-layer(seeing Fig.\ref{A-layer}), other valuable image pixels get more attention by increasing the weights of high-quality and lower the low-quality pixels. As a consequence, this structure can help  CANet  to  rebuild the corrupted information from low-quality images.

\subsection{Channel-wise Attention}

The channel-wise attention contains several steps(seeing Fig.\ref{PAandCA}). Firstly, feature maps are fed into the channel-wise global average pooling layer:
\begin{align}
y_c=\frac{1}{H\times W}\sum_{i=1}^{H}\sum_{j=1}^{W}x_c(i,j),
\end{align}
where $x_c(i,j)$ is the feature value at position $(i, j)$ in the $c$-th channel of feature map $X$. The global average pooling operator collects the channel-wise global information and change the feature map from $H\times W \times C$ to $1\times 1 \times C$. $y_c$ is then passed into two convolution layers, one PRelu and sigmoid activation operator, which gets the weight of the channel-wise attention. Furthermore, we have  
\begin{align}
r_c=\alpha(Conv_2(\delta(Conv_1(y_c)))),
\end{align}
where $Conv_1$  reduces the channel of $y_c$ and $Conv_2$   upsamples the channel. $\delta$ and $\alpha$ are the PRelu  and sigmoid operators.
Finally, multiplying  the input $x_c$ and $r_c$ with element-wise, we  get
\begin{align}
f_c=r_c*x_c.
\end{align}
The whole process is showed as Fig.\ref{A-layer}. By adding the channel attention layer after the pixel attention, instead of only focusing on pixel weights, A-layer can balance the issue of useful and corrupted information in different channels.

\subsection{Loss Function}

For low-level vision task, several different loss functions like $L_2$ loss\cite{mao2016image}, $L_1$ loss\cite{zhang2018residual}, perceptual loss and adversarial loss\cite{ledig2017photo} have been used. In order to comparing with other methods fairly, especilly RNAN\cite{zhang2019residual} and IDCN\cite{Zheng_2019}, $L_2$ loss function is choosed:
\begin{align}
L(\theta) = \frac{1}{N}\sum_{i=1}^{N}\Vert O_{HQ}-I_{HQ} \Vert^2,
\end{align}
where $\theta$ is the parameter of CANet. $O_{HQ}$ stands for the output of the network, where $I_{HQ}$ is the ground-truth and $\Vert\Vert^2$ denotes $l_2$ norm.

\subsection{Implementation Detials}
 In this paper, hyperparameters are set as follows. CANet contains 5 A-blocks, and each block is comprised by 6 A-layers. The kernel size for convolutional layers is generally $3\times 3$. For pixel-wise  and channel-wise attention, the $1 \times 1$ convolution is employed. The number of channels for the output of A-layers is fixed as 64.

\section{Experiments}

\subsection{Datasets and Metrics}
We conduct experiments focusing on tasks of image denoising and JPEG compression artifacts reduction.  DIV2K\cite{timofte2017ntire} is employed for training our models. To keep the consistency with the original work\cite{zhang2019residual} \cite{Zheng_2019}, for compress artifacts reduction, all the 900 DIV2K images (including 800 DIV2K training data and 100 DIV2K validation data)are used for training, while the 100 validation images from BSDS500 are employed for validation. As for image denoising, we choose the 800 DIV2K  data for training and Kodak24 dataset for validation, and extract 16 patches with size of $48 \times 48$ from each inputs. Adam\cite{kingma2014adam} is used as the optimizer. The learning rate is set to $10^{-4}$ and all models are trained with  1080Ti GPU and RTX2080Ti. In addition, common dataset is employed in every task for testing and  peak signal-to-noise ratio(PSNR) and structural similarity index(SSIM)as the  evaluation metric.

\subsection{Compression Artifacts Reduction}
For image compression artifacts problem, CANet is compared  with other state-of-the-art algorithms  including ARCNN\cite{Dong2016Compression}, RDN\cite{zhang2020residual}, DPW-SDNET\cite{chen2018dpwsdnet}, MemNet\cite{tai2017memnet}, MWCNN\cite{liu2018multilevel},S-NET\cite{zheng2018s} and IDCN\cite{Zheng_2019}. Part of the test data is consistent with the IDCN. Two latest state-of-the-art methods(RIDNet\cite{anwar2019real}, RNAN\cite{zhang2019residual}) are also included in Table.\ref{CAR}. Test data includes 29 images from LIVE1, 200 images from BSDS500(B200), and 143 images from WIN143.

\begin{figure}[htbp]
\centering
\subfigure[original]{
\includegraphics[width=1.0 in]{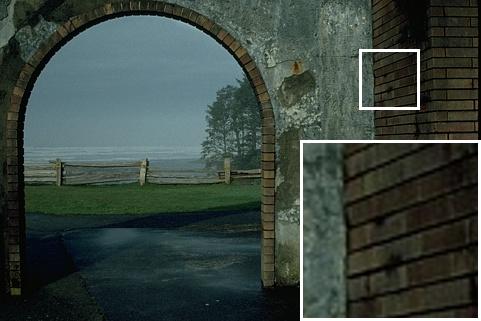}
}	
\subfigure[jpeg]{
\includegraphics[width=1.0 in]{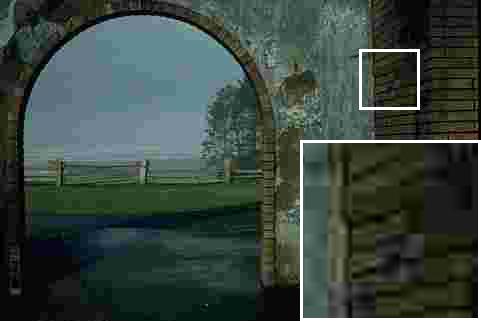}
}
\subfigure[ARCNN]{
\includegraphics[width=1.0 in]{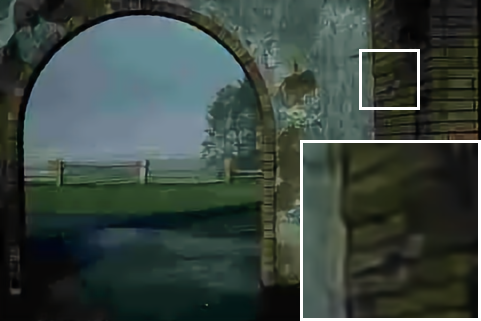}
}	
\subfigure[S-NET]{
\includegraphics[width=1.0 in]{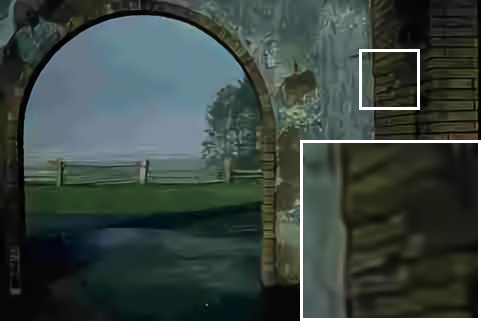}
}
\subfigure[MemNet]{
\includegraphics[width=1.0 in]{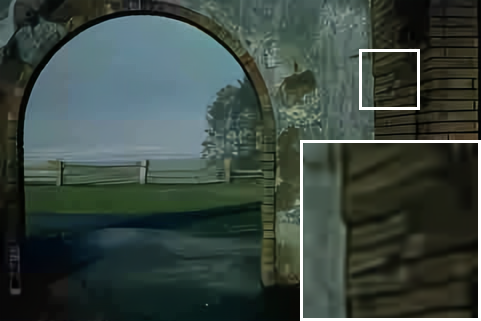}
}
\subfigure[RNAN]{
\includegraphics[width=1.0 in]{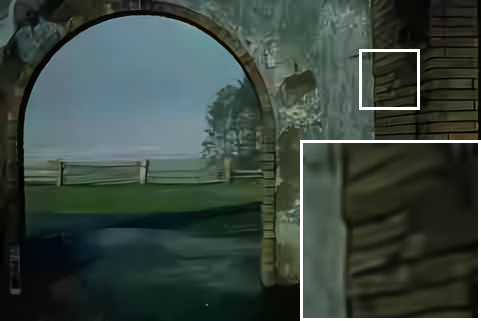}
}
\subfigure[RIDNET]{
\includegraphics[width=1.0 in]{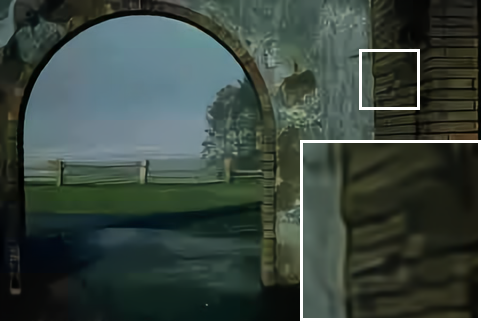}
}
\subfigure[GLC-Net]{
\includegraphics[width=1.0 in]{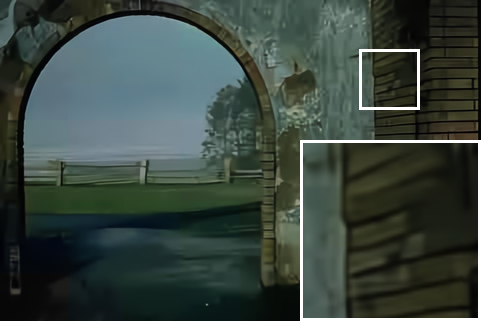}
}
\caption{ '5096' from BSDS500, subjective evaluation with jpeg quality q=10.}
\label{jpeg1}
\end{figure}

As Table.\ref{CAR} shows, for the compression quality, our method gets the best  results in all datasets.  Not only that, the more complex the compressed images are, the better our  network  performs, we can see  when the quality=10, the  PSNR  of  CANet  is only 0.04dB higher then  IDCN  for  LIVE1, while for more complex B200 and WIN143,  it turns to be   0.70dB  and  0.36dB  respectively,   which also indicates  that  CANet  has  more powerful representation ability. In order to reduce the compression artifacts, some methods like MWCNN and IDCN tend to use the DCT domain, however, it usually has less generality for other low-level vision tasks. Thus, we try to adopt the pixel domain to create a  more general framework  being able to handle other image tasks. In addition,  experiments  show  that  CANet  outperforms  RIDNet  and  RNAN  because  of using both pixel-wise and channel-wise attention. Although RIDNet used the channel-wise attention layer, it was not enough to deal with the CAR and resulted the limit performance, while  RNAN considered the spatial information by using non-local mechanism[22], its performance was still limited by the high computational costs. Besides, the task of maintaining lightweight(see Table.\ref{paramters}) makes CANet more powerful. In CAR task, LR images loss a lot of high-frequency information due to the lossy compression algorithm. In practice, visual results for compression artifacts reduction are presented in Fig.\ref{jpeg1}. According to image ’5096’ from BSDS500, most methods produce the blurring  artifacts  or  the  out  of  shape  image,  but  our  method  generates  more clear textures.
\setlength{\tabcolsep}{6pt}
\begin{table}
\setlength{\belowcaptionskip}{-0.1cm} 
  \centering
  \caption{The compression artifacts reduction for RGB images on LIVE1, BSDS500, WIN143. Average PSNR and SSIM as metrics on quality 10 and 20. Best results are red.}
    \begin{tabular}{c|c|c|c|c|c|c|c}
   	\hline
\hline
    \multicolumn{1}{r|}{\multirow{2}[2]{*}{quality}} & \multicolumn{1}{c|}{\multirow{2}[2]{*}{Methods}} & \multicolumn{2}{c|}{LIVE1} & \multicolumn{2}{c|}{B200} & \multicolumn{2}{c}{WIN143} \\
          &       & \multicolumn{1}{l}{PSNR} & \multicolumn{1}{l|}{SSIM} & \multicolumn{1}{l}{PSNR} & \multicolumn{1}{l|}{SSIM} & \multicolumn{1}{l}{PSNR} & \multicolumn{1}{l}{SSIM} \\  [1.5pt] \hline

    \multirow{9}[2]{*}{20} & JPEG  & 28.06 & 0.8409 & 28.2  & 0.8483 & 29.47 & 0.844 \\ [1.5pt]
          & ARCNN & 29.23 & 0.8659 & 29.36 & 0.8665 & 30.82 & 0.8776 \\ [1.5pt]
          & DPW-SDNET & 29.59 & 0.8744 & 29.67 & 0.8752 & 31.28 & 0.8866 \\ [1.5pt]
          & MemNet  & 29.76 & 0.8770 & 29.80  & 0.8776 & 31.47 & 0.8904 \\ [1.5pt]
          & MWCNN & 29.80  & 0.8769 & 29.85 & 0.8789 & 31.55 & 0.8916 \\ [1.5pt]
          & RDN   & 29.84 & 0.8778 & 29.85 & 0.8779 & 31.54 & 0.8912 \\ [1.5pt]
          & S-Net & 29.81 & 0.8781 & 29.86 & 0.8782 & 31.47 & 0.8904 \\ [1.5pt]
		& RIDNet  & 29.62& 0.8785 &29.72 &0.8829  & 31.70 & 0.8830\\ [1.5pt]
		& RNAN  & 30.07  & 0.8852 & 30.11& 0.8889 & 32.12 & 0.8888\\ [1.5pt]
          & IDCN  & 30.04 & 0.8816 & 30.07 & 0.8816 & 31.82 & \textcolor[rgb]{1, 0, 0}{0.8964} \\ [1.5pt]
          & CANet   &  \textcolor[rgb]{ 1,  0,  0}{30.08}      & \textcolor [rgb]{ 1,  0,  0}{0.8852}       & \textcolor[rgb]{ 1,  0,  0}{30.15}      & \textcolor[rgb]{ 1,  0,  0}{0.8897}      &  \textcolor[rgb]{ 1,  0,  0}{32.18}     & 0.8900 \\ [1.5pt] \hline
    \multirow{9}[2]{*}{10} & JPEG  & 25.69 & 0.7592 & 25.83 & 0.7584 & 27.08 & 0.7684 \\ [1.5pt]
          & ARCNN & 26.91 & 0.7946 & 27.02 & 0.793 & 28.46 & 0.8207 \\ [1.5pt]
          & DPW-SDNET & 27.26 & 0.8036 & 27.39 & 0.8027 & 29.03 & 0.8326 \\ [1.5pt]
          & MemNet  & 27.33 & 0.8100  & 27.46 & 0.8086 & 29.04 & 0.8380 \\ [1.5pt]
          & MWCNN & 27.45 & 0.8083 & 27.52 & 0.8069 & 29.25 & 0.8375 \\ [1.5pt]
          & RDN   & 27.47 & 0.8116 & 27.53 & 0.8096 & 29.19 & 0.8395 \\ [1.5pt]
          & S-Net & 27.35 & 0.8090 & 27.42 & 0.8066 & 28.95 & 0.8349 \\ [1.5pt]
		 & RIDNet  & 27.35 & 0.8122 & 27.48 &0.8149 & 29.47 & 0.8336 \\ [1.5pt]
		 & RNAN  & 27.40 & 0.8117  & 27.51 & 0.8143  & 29.48 & 0.8318 \\ [1.5pt]
          & IDCN  & 27.63 & 0.8161 & 27.69 & 0.8136 &  29.45& 0.8467 \\ [1.5pt]
		  
          & CANet   &  \textcolor[rgb]{ 1,  0,  0}{27.67}     &  \textcolor[rgb]{ 1,  0,  0}{0.8192}       & \textcolor[rgb]{ 1,  0,  0}{27.77}    &  \textcolor[rgb]{ 1,  0,  0}{0.8221}      & \textcolor[rgb]{ 1,  0,  0}{30.15}      & \textcolor[rgb]{ 1,  0,  0}{0.8897} \\ \hline \hline
    \end{tabular}%
  \label{CAR}%
\end{table}%

\subsection{Image Denoising}

In the image denoising task, some of the results consistently with zhang et al\cite{zhang2019residual} are kept, and the Kodak24 is used for validation dataset and test in Urban100 and CBSD68. For each dataset, we had different AWGN noises level(e.g., 10, 30, 50) to be tested.  The comparison methods include: ARCNN\cite{Dong2016Compression},  CBM3D\cite{dabov2007image},  TNRD\cite{Chen2015Trainable}, RED\cite{mao2016image},  DnCNN\cite{zhang2017beyond},  MemNet\cite{tai2017memnet}, IRCNN\cite{zhang2017learning},  FFDNet\cite{zhang2018ffdnet},  
RIDNet\cite{anwar2019real} and RNAN\cite{zhang2019residual}. 

\begin{figure}
\centering
\subfigure[original]{
\includegraphics[width=1.0 in]{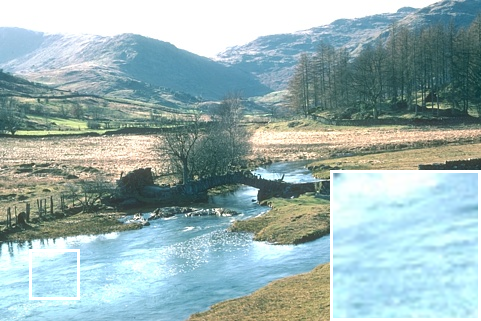}
}	
\subfigure[noise]{
\includegraphics[width=1.0 in]{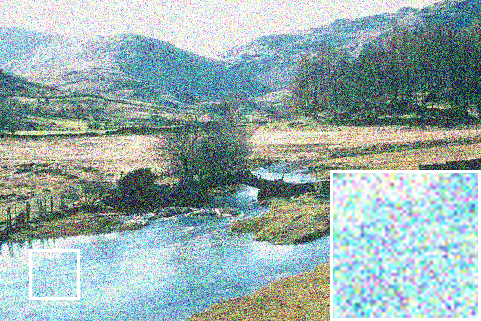}
}
\subfigure[ARCNN]{
\includegraphics[width=1.0 in]{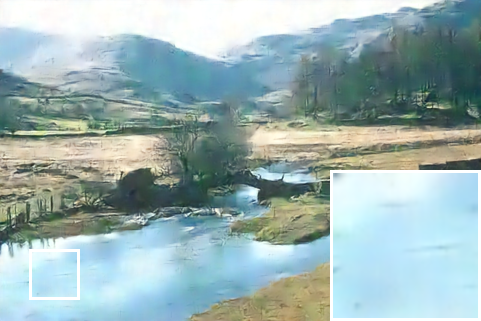}
}	
\subfigure[MemNet]{
\includegraphics[width=1.0  in]{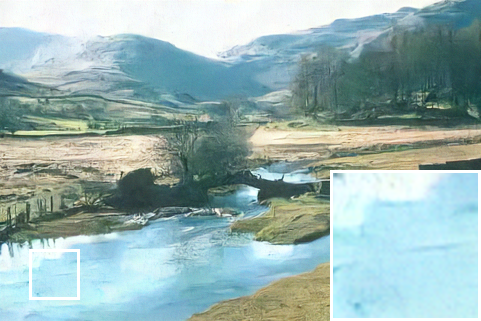}
}
\subfigure[DnCNN]{
\includegraphics[width=1.0  in]{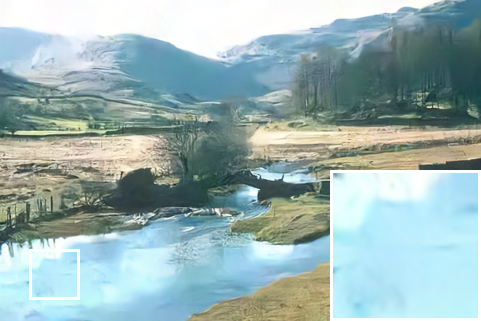}
}
\subfigure[RIDNET]{
\includegraphics[width=1.0 in]{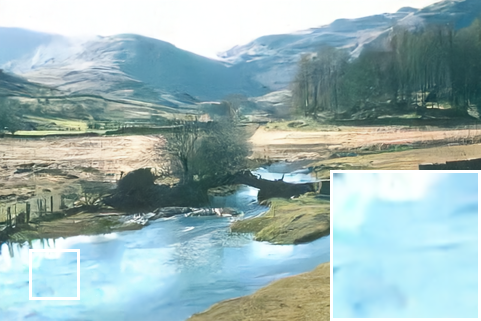}
}
\subfigure[RNAN]{
\includegraphics[width=1.0  in]{CBSD68/DnCNN+N50/241048_x1_SR_new.png}
}
\subfigure[CANet]{
\includegraphics[width=1.0  in]{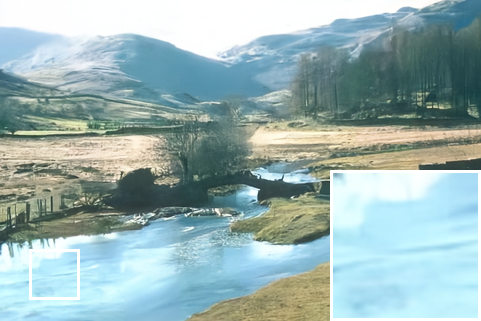}
}
\caption{'241048' from CBSD68, subjective evaluation with noise $\sigma$=50.}
\label{denoising1}
\end{figure}

\begin{figure}
\centering
\subfigure[original]{
\includegraphics[width=1.0  in]{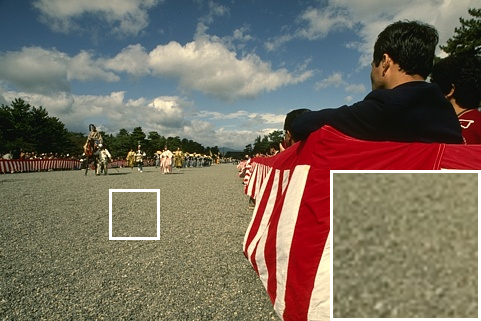}
}	
\subfigure[jpeg]{
\includegraphics[width=1.0   in]{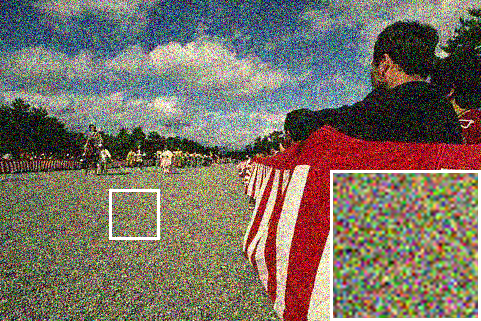}
}
\subfigure[ARCNN]{
\includegraphics[width=1.0 in]{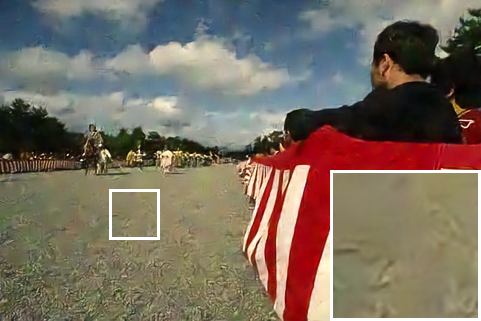}
}	
\subfigure[MemNet]{
\includegraphics[width=1.0   in]{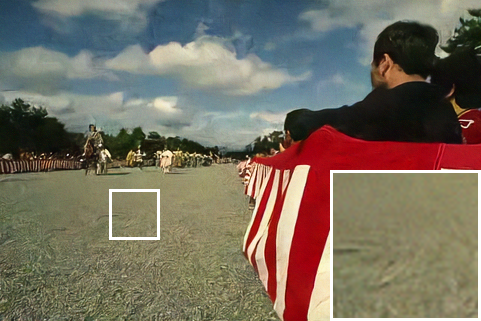}
}
\subfigure[DnCNN]{
\includegraphics[width=1.0 in]{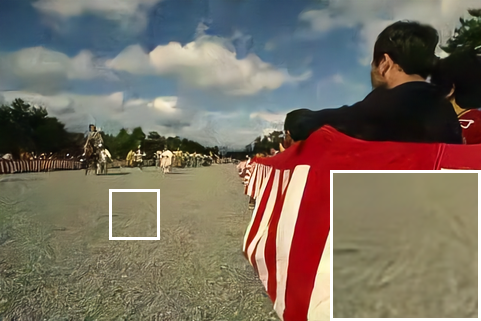}
}
\subfigure[RIDNET]{
\includegraphics[width=1.0 in]{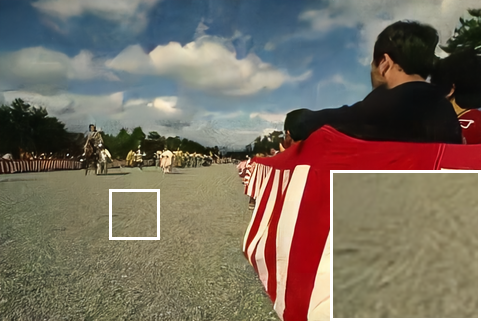}
}
\subfigure[RNAN]{
\includegraphics[width=1.0  in]{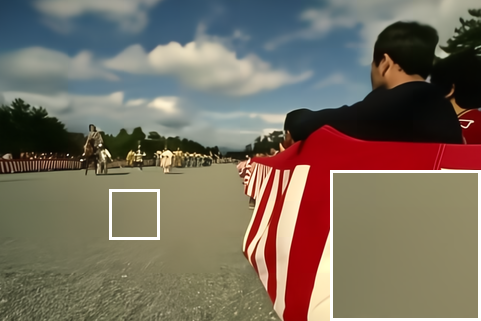}
}
\subfigure[GLC-Net]{
\includegraphics[width=1.0 in]{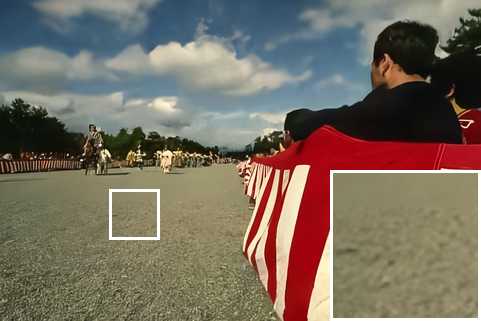}
}
\caption{'145086' from CBSD68, subjective evaluation with noise $\sigma$=50. }
\label{denoising2}
\end{figure}

All results are showed in Table.\ref{denoising_table}. Not surprisingly, CANet has the best results on all the datasets under different noise levels. RNAN  also has a superior performance, but CANet has less than half of parameters than RNAN (see Table.\ref{paramters}). CANet is 3.26M  far less  than 7.5M RNAN. For image denoising task, it's important and difficult to reconstruct the high-frequency information while removing the noise. The visual results are also given in Fig.\ref{denoising1} and Fig.\ref{denoising2}. Taking the '241048' from CBSD68 as an  example, We give the river AWGN noise($\sigma$=50) and  the visual comparisons. Although most methods have obvious effects in terms of denoising,  they nearly suffer from the serious loss of detail(ripples in river). But  CANet can keep the more detailed contents than other methods. Seeing  the '145086' from CBSD68. While target is removing the noise and keeping the texture of grass, the figure shows that RNAN loses almost all texture information,  but CANet better preserves the  texture of grass.

\setlength{\tabcolsep}{12.5pt}
\begin{table}
  \centering
  \caption{RGB image denosing quantitative results on Urban100 and CBSD68. The best results are set to red.}
    \begin{tabular}{cc|c|c|c|c|c|c}
    
	\hline
	\hline
	\multicolumn{2}{c|}{\multirow{2}[2]{*}{Method}} & \multicolumn{3}{c|}{CBSD68} & \multicolumn{3}{c}{Urban100} \\ [1.5pt]
    \multicolumn{2}{c|}{} & \multicolumn{1}{c}{10} & \multicolumn{1}{c}{30} & 50    & \multicolumn{1}{c}{10} & \multicolumn{1}{c}{30} & 50     \\ [1.5pt] \hline
    
    \multicolumn{2}{c|}{CBM3D}& 35.91 & 29.73 & 27.38  & 36    & 30.36 & 27.94  \\ [1.5pt]
    \multicolumn{2}{c|}{TNRD}& 33.36 & 27.64 & 25.96  & 33.6  & 27.4  & 25.52  \\ [1.5pt]
    \multicolumn{2}{c|}{ARCNN}& 35.91 & 29.73 & 27.27  & 34.97    & 29.22 & 26.64  \\ [1.5pt]
    \multicolumn{2}{c|}{RED} & 33.89 & 28.46 & 26.35  & 34.59 & 29.02 & 26.4   \\ [1.5pt]
    \multicolumn{2}{c|}{DnCNN} & 36.31 & 30.4  & 28.01 & 36.21 & 30.28 & 28.16  \\ [1.5pt]
    \multicolumn{2}{c|}{MemNet}& \multicolumn{1}{l|}{N/A} & 28.39 & 26.33 & \multicolumn{1}{c|}{N/A} & 28.93 & 26.53  \\ [1.5pt]
    \multicolumn{2}{c|}{IRCNN} & 36.06 & 30.22 & 27.86 & 35.81 & 30.28 & 27.69   \\ [1.5pt]
    \multicolumn{2}{c|}{FFDNet} & 36.14 & 30.31 & 27.96 & 35.77 & 30.53 & 28.05  \\ [1.5pt]
    \multicolumn{2}{c|}{RIDNet} & 36.21 & 30.35  &  28.09 & 36.08 & 30.21  & 28.42  \\ [1.5pt]
    \multicolumn{2}{c|}{RNAN} & 36.43 & 30.63 & 28.27 & 36.59 & 31.50& 29.08  \\ [1.5pt]
	
    \multicolumn{2}{c|}{CANet} & \textcolor[rgb]{ 1,  0,  0}{36.45} & \textcolor[rgb]{ 1,  0,  0}{30.88}   & \textcolor[rgb]{ 1,  0,  0}{28.29	}  & \textcolor[rgb]{ 1,  0,  0}{36.62} & \textcolor[rgb]{ 1,  0,  0}{31.51}     &   \textcolor[rgb]{ 1,  0,  0}{29.12}       \\[1.5pt] \hline \hline    
    \end{tabular}%
  \label{denoising_table}%
\end{table}%
	
\setlength{\tabcolsep}{6pt}
\begin{table}
  \centering
  \caption{The comparison of the number of parameters for different methods in $\sigma$=50 Urban100.}
    \begin{tabular}{l|c|c|c|c|c|c}
	\hline
	\hline
    Methods & ARCNN     & MemNet   & DnCNN    & RIDNET     & RNAN    & GLCNet \\ [1.5pt] \hline
    PSNR  &  26.64     &  26.53     &  28.16     &   28.42    &  29.08     & 29.12 \\ [1.5pt]
    params & 0.12M       & 2.91M      &  0.55M     &  1.50M    & 7.5M      & 3.26M \\ [1.5pt]
	\hline
	\hline
    \end{tabular}%
  \label{paramters}%
\end{table}%

\subsection{Ablation Study}

In this section, the relevant ablation study is conducted, all results are showed in Table.\ref{block number} and Table.\ref{Components}. \textbf{First}, we compare the effect of different A-block numbers on the PSNR of the results(seeing Table.\ref{block number}). In the case of using only one block(one block have just 0.65M parameters), we can achieve 36.30dB. which has exceeded most methods for 0.09dB higher than RIDNet and 0.4dB higher than ARCNN.The 5 blocks achieves a performance improvement of 0.15dB over 1 block. But more blocks bring more  computational cost. So in practice, to keep our network lightweight, we choose the 5 blocks for constituting CANet. \textbf{Second}, we compare the performance of  different components on the PSNR of the results.   As the Table.\ref{Components} shows, we choose the best results for every components with epoch 500. The other configures is the same as CANet implementation details. When the concatenation operation is only introduced,  the experiment shows that it is critical to improve our algorithm performance 0.02dB than element-wise operator. And then, the feature selection operator is also added the concatenation network. We get another 0.01dB improvement in PSNR. For the CANet, we get the best results 37.18dB, which shows that every components we propose are important for the final performance.

\setlength{\tabcolsep}{13pt}
\begin{table}

  \centering
  \caption{The ablation study of different block numbers. PSNR values are based on CBSD68($\sigma$=10)}
    \begin{tabular}{c|c|c|c|c|c}
	\hline
	\hline
    Block Number & 1     & 2     & 3     & 4     & 5      \\ [1.5pt] \hline
    PSNR  &   36.30    &  36.38     &   36.41    &        36.44    &  36.45 \\ [1.5pt]
	\hline
	\hline
    \end{tabular}%
  \label{block number}%
\end{table}%

\setlength{\tabcolsep}{17pt}
\begin{table}
\setlength{\belowcaptionskip}{-1cm} 
  \centering
  \caption{The ablation study of different components.PSNR values are based on  Kodak24 ($\sigma$=10, epoch=500) }
    \begin{tabular}{c|c|c|c|c}
	\hline
	\hline
    element-wise  &    $\surd$   &  $\times$      &    $\times$     &  $\times$        \\ [1.5pt]
    concatenation   &  $\times$     &   $\surd$     &   $\surd$    &  $\surd$       \\ [1.5pt]
    feature-selection &   $\times$     & $\times$       &   $\surd$    &   $\surd$      \\ [1.5pt]
    feature-attention  &   $\times$     &  $\times$      &   $\times$     &  $\surd$       \\ [1.5pt] \hline
    PSNR & 37.11     & 37.13     & 37.14     & 37.18  \\ [1.5pt] 
   
	\hline
	\hline
    \end{tabular}%
  \label{Components}%
\end{table}%

\section{Conclusions}
In  this  paper,  we  present  a general framework for low-level vision tasks including image compression artifacts reduction and image denoising. Under this framework, CANet is built with several residual attention blocks for image restoration, which has the powerful ability extracting features of images. Next, we use the feature selection and concatenation operation to connect blocks, the result of which also be used as universal  means for low-level vision tasks like image compress artifacts remove, image denoising.  Experiments show that our method outperforms several state-of-the-art methods on benchmark datasets.  However, in fact, the  restoration  of  visual  details  and  texture  information  is  still difficult under low-level vision tasks. And none of the existing methods can solve this problem perfectly. In the future work, we plan to recover more details and texture information for image restoration and extend our framework for other tasks like super-resolution, image dehazing, image deblurring.

\clearpage
%
%
\bibliographystyle{splncs04}
\bibliography{egbib}
\end{document}